# Dynamic Pricing of Electric Vehicle Charging Station Alliances Under Information Asymmetry

Zeyu Liu, Yun Zhou, *Member*, *IEEE*, Donghan Feng, *Senior Member*, *IEEE*, Shaolun Xu, Yin Yi, Hengjie Li, Haojing Wang

*Abstract*—Due to the centralization of charging stations (CSs), CSs are organized as charging station alliances (CSAs) in the commercial competition. Under this situation, this paper studies the profit-oriented dynamic pricing strategy of CSAs. As the practicability basis, a privacy-protected bidirectional real-time information interaction framework is designed, under which the status of EVs is utilized as the reference for pricing, and the prices of CSs are the reference for charging decisions. Based on this framework, the decision-making models of EVs and CSs are established, in which the uncertainty caused by the information asymmetry between EVs and CSs and the bounded rationality of EV users are integrated. To solve the pricing decision model, the evolutionary game theory is adopted to describe the dynamic pricing game among CSAs, the equilibrium of which gives the optimal pricing strategy. Finally, the case study results in a real urban area in Shanghai, China verifies the practicability of the framework and the effectiveness of the dynamic pricing strategy.

## I. Introduction

### A. Motivation

THE advantages of electric vehicles (EVs) have led to a rapid increase in their numbers in recent years, prompting optimism among researchers about their future growth [1]. However, this surge in EV adoption has also resulted in fierce competition, causing a significant concentration of charging facilities. In China, as depicted in Fig. 1, by October 2023, approximately 65.5% of public charging stations were controlled by the top 5 charging service providers (CSPs).

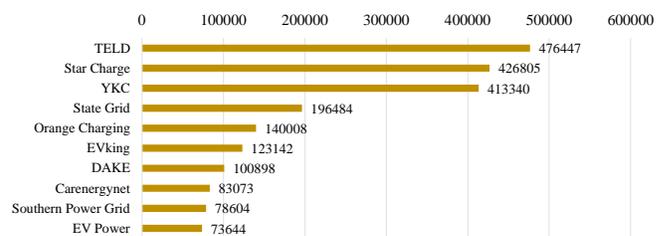

Fig. 1. Public charging pile numbers of the top 10 CSPs in China by October 2023.

The concentration of charging infrastructure leads to intensified competition among a limited number of CSPs. It is presumed that charging stations (CSs) under the same CSP naturally form charging station alliances (CSAs), collaborating internally while competing against other CSAs to maximize overall profit. Apart from geographical locations, pricing stands out as the most impactful method for CSs to entice EVs. Notably, the pricing strategy employed by CSAs differs significantly from that of an individual CS. As an illustrative example in Fig. 2 demonstrates, assuming equal distances to CS1 and CS2 for a perfectly rational EV user, an independent CS1 would ideally lower its price to 1.4 ¥/kWh to attract the EV and earn 14 ¥. However, if CS1 and CS2 are part of the same CSA, CS1 can maintain its price while the CSA collectively earns 15 ¥.

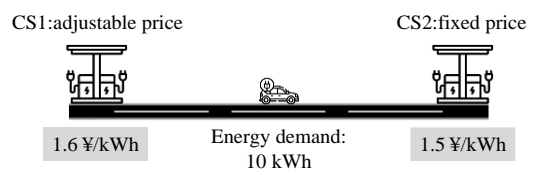

Fig. 2. An illustrative example explaining the strategy difference.

However, currently, limited attention has been devoted to examining the dynamic pricing within CSAs. This prompts us to delve into an analysis of dynamic pricing among CSAs.

Recent efforts have begun to explore dynamic pricing strategies. Notably, Spirii has enabled charging site operators to transition from fixed rates to adaptable pricing based on real-time market shifts [2]. Autopay has also embraced dynamic pricing in parking and charging management [3]. Although there are no reported instances of dynamic pricing adoption in China, governmental policies encourage such innovative endeavors. A June 2023 State Council document emphasized the need to "promote the development of new technologies, new formats, and new models" and to "enhance technological research in information sharing and unified settlement systems" [4]. In essence, the concept of dynamic pricing for EV charging enjoys widespread acceptance, with ongoing implementations validating the feasibility of this research.

*B. Literature Review*

Many studies have explored charging service pricing for diverse purposes. For instance, reference [5] introduces a coordinated dynamic pricing approach aimed at reducing load overlaps. Reference [6] proposes a pricing mechanism that assigns additional fees resulting from EV charging during peak demand hours, thereby encouraging off-peak charging. While these pricing methods benefit the grid, they often overlook the financial objectives of CSs. Other pricing objectives studied in existing research encompass minimizing voltage offsets and network losses [7], expediting CS cost recovery [8], balancing charging site utilization [9], and promoting private charging pile sharing to alleviate facility shortages [10][11].

However, some researchers advocate viewing CSs as profit-maximizing economic entities. Reference [12] delves into financially driven competition among independent CSs, modeled as a super-modular game, but without considering CS alliances. Reference [13] establishes a trilevel optimization model that accounts for distribution locational marginal prices, yet overlooks EV reactions. Besides the above studies, references [14][15][16] also focus on pricing for independent CSs aiming at profit maximization. Reference [17] addresses a multi-objective dynamic pricing problem for a CSP operating several CSs, organizing CSs as a CSA, but overlooking competition among different CSAs.

The prevailing approach in these studies is to model charging service pricing as a Stackelberg game, involving a decision leader (CS) and followers (EVs) [10][14][15][16][18]. Machine learning methods are also being explored for dynamic CS pricing [19][20][21][22][23], offering an alternative methodology provided adequate historical data is available. However, these methods demand reliable data, raising concerns regarding their applicability across different contexts. A common assumption in these studies is that EV users are perfectly rational decision-makers whose response to charging prices is predictable. However, in reality, EV users possess bounded rationality, amplifying the information asymmetry between EVs and CSs. This challenges the practicality of the Stackelberg game methodology.

Information asymmetry remains a significant concern in the dynamic pricing of CSs. Alongside decision preferences, the lack of information regarding EV arrival times and charging demands poses a major challenge for CSs [24]. Most studies model EV arrivals probabilistically or make deterministic assumptions about EV availability. This asymmetry hampers efficient decision-making for EVs, as exemplified in [25].

Efforts have been made to mitigate this asymmetry through communication measures. In [26], an approach is described where EVs send charging reservation requests and real-time status updates to a cloud server, allowing them to accept or reject schedule feedback. Additionally, [27] explores the use of blockchain technology and 6G communication networks. Privacy preservation for CSs is considered in a decentralized pricing method proposed in [28]. In summary, few studies have leveraged real-time bidirectional information exchange to dynamically adjust EV status for pricing due to concerns about practicability. However, according to our conversations with EV drivers, they are willing to utilize charging or map applications to make optimal choices while authorizing the collection of non-privacy data like location information.

To conclude, while CSAs represent the most realistic competitive CS entities, existing studies largely overlook discussions regarding the dynamic pricing of CSAs. Moreover, these studies inadequately address information asymmetry, underscoring the underutilization of real-time information communication in charging pricing to enhance timeliness and accuracy.

*C. Contributions*

Diverging from existing literature, our study introduces a dynamic pricing strategy for CSAs within a bidirectional communication framework. The main contributions of this research can be succinctly outlined in the following 4 points:

1. Our paper pioneers a dynamic pricing strategy for CSs in the form of competitive alliances. This innovative approach offers a fresh perspective for the analysis of CS strategies.

2. A two-stage evolutionary game is devised to determine optimal pricing decisions for CSAs within an expansive action space. Our methodology can identify near-optimal strategies efficiently, within a reasonable timeframe.

3. The information asymmetry between EVs and CSs is adeptly integrated in the proposed model. The stochastic modeling of the time sensitivity coefficient under the communication framework handles asymmetric information.

4. The bounded rationality of EV users is given full consideration. The value function model, mental account theory, and Logit model are applied to realistically depict the decision-making process of EV users.

These contributions collectively enhance the understanding and approach toward dynamic pricing strategies in the form of CSAs, offering a comprehensive framework that encapsulates various critical factors influencing decision-making processes in this domain.

*D. Structure*

The subsequent sections are structured as follows:

Section II: Explores the foundation of dynamic pricing through real-time bidirectional communication.

Section III: Introduces the model basis, detailing the decision-making models of both EVs and CSs.

Section IV: Presents the solution methodology of the pricing problem. The pricing is modeled as a two-stage evolutionary game, whose equilibrium gives the optimal strategies.

Section V: Conducts a case study within a specific area in Shanghai, China, to validate the practicality and efficacy of the proposed framework and strategy.

Section VI: Concludes the paper and provides a glimpse into potential avenues for future research.

## II. REAL-TIME COMMUNICATION FRAMEWORK

Upon analysis, it is evident that real-time bidirectional communication holds practical significance in EV charging. This section delves into the operational mechanics of the real-time communication framework.

### A. Information mode

The information pertaining to EVs and CSs can be categorized into public and private information. Public information is shared in real time through communication platforms like cellular networks or the Internet of Vehicles (IoV). Private information is accessible only to its respective owner. The communication platforms themselves don't generate new data, but embedded online map services can offer derivative outcomes based on public information.

(1) Information of EVs

Let $I$ denote the set of EVs. The public information of EV $i \in I$ comprises real-time position, real-time State of Charge (SOC), target SOC, battery capacity, and time limit. The real-time position and SOC are automatically obtained, while the user provides the remaining data; default values may substitute if the EV user declines or fails to provide this information. Private information includes the time sensitivity coefficient, which represents the importance of travel time in charging decisions (further details are explored in section III). Table I provides a summary of this information.

TABLE I
INFORMATION OF EVs

| Privacy | Content | Symbol | Meaning | Default |
|---|---|---|---|---|
| Public | Real-time position | $(x_{EV,i}, y_{EV,i})$ | Real-time longitude and latitude | - |
| | Real-time SOC | $Soc_i$ | Real-time SOC | - |
| | Target SOC | $Soc_i^{target}$ | The SOC wanted when leaving CS | 0.8 |
| | Battery capacity | $C_i$ | Battery capacity | 60 kWh |
| | Time limit | $t_i^{limit}$ | Available charging time before leaving | Minimal time to reach $Soc_i^{target}$ |
| Private | Time sensitivity coefficient | $\theta_i$ | The weight of travel time in decision | - |

(2) Information of CSs

Let $J$ denote the set of CSs. The public information of CS $j \in J$ includes the CS's location, the real-time charging price, and the expected waiting time. The private information of CSs includes the perceived time sensitivity coefficient distribution (also discussed in section III). Table II gives a summary.

TABLE II
INFORMATION OF CSs

| Privacy | Content | Symbol | Meaning | Default |
|---|---|---|---|---|
| Public | Location | $(x_{CS,j}, y_{CS,j})$ | The longitude and latitude of CS | - |
| | Real-time charging price | $p_j$ | The price of 1 kWh of electricity | - |
| | Expected waiting time | $T_j^{wait}$ | The waiting time of EVs after arriving at the CS | 0 |
| Private | Perceived time sensitivity coefficient distribution | $\Theta$ | The distribution of EV fleet's time sensitivity coefficients from CS's point of view | - |

### B. Behavior mode

CSs periodically make pricing decisions according to the existing charging requests and broadcast the real-time charging price in the next period. EV users propose charging requests at any random time and broadcast the public information simultaneously. Meanwhile, EV users browse the information of nearby CSs, make a decision, and drive to the target station or delay the charging. Before an EV is attached to a charging pile, the user can still change the decision.

Fig. 3 visually illustrates the time sequence and the information interaction relationship.

## III. DECISION-MAKING MODEL

### A. EVs' decision-making model

The main factors in EVs' decision-making can be summarized as the expense cost and the time cost. Due to the limited decision time and incomplete information, users can hardly calculate the precise utilities of each choice. Instead, they will pursue "satisfaction" criteria in their decisions. Therefore, we use the cumulative prospect theory to describe EV users' decision-making process under bounded rationality.

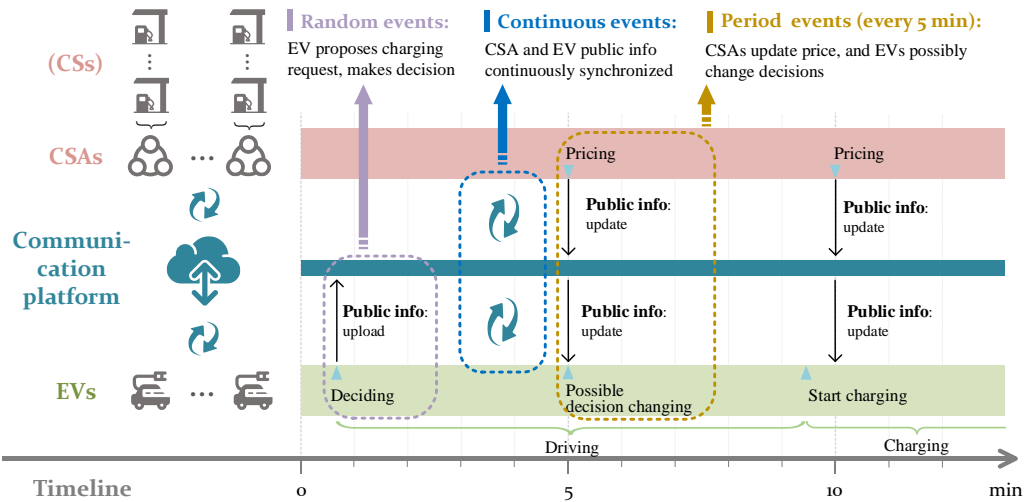

Fig. 3. The real-time communication framework.

*1) Expense cost*

The expense cost of EV *i* if charged at CS *j*, denoted by $M_{i,j}$, is the product of the charging price and the energy demand.

$$M_{i,j} = p_j E_{i,j} \quad (1)$$

where $p_j$ is the charging price of CS *j* at the decision moment, and $E_{i,j}$ is the energy demand of EV *i*. $E_{i,j}$ is expressed as

$$E_{i,j} = \min\left\{E_i^{\text{target}}, E_{i,j}^{\max}\right\} \quad (2)$$

where $E_i^{\text{target}}$ is the energy demand to reach the target SOC, and $E_{i,j}^{\max}$ is the maximal possible charging energy at CS *j*.

$$E_i^{\text{target}} = \left(Soc_i^{\text{target}} - Soc_i\right)C_i \quad (3)$$

$$E_{i,j}^{\max} = q_{i,j}^{\text{ch,max}} t_i^{\text{limit}} \quad (4)$$

where $Soc_i^{\text{target}}$ denotes the target SOC, set by the EV user, with a default value of 0.8; $Soc_i$ is the real-time SOC, $C_i$ is the battery capacity, $t_i^{\text{limit}}$ is the time limit, $q_{i,j}^{\text{ch,max}}$ is the maximal charging power of EV *i* at CS *j*.

*2) Time cost*

The time cost $T_{i,j}$ is the sum of driving time and waiting time. Denote the driving time of EV *i* to CS *j* as $T_{i,j}^{\text{drive}}$, and the estimated waiting time after arriving at CS *j* as $T_j^{\text{wait}}$, then

$$T_{i,j} = T_{i,j}^{\text{drive}} + T_j^{\text{wait}} \quad (5)$$

In most current research, driving time is typically calculated using road network modeling and route planning algorithms. However, these highly simplified simulations of real traffic within their road networks lack the capacity to factor in occurrences like temporary traffic control or accidents. Leveraging professional online map services, such as Google Maps, offers access to robust data sources and highly optimized route planning technologies, ensuring both speed and accuracy. $T_{i,j}^{\text{drive}}$ can be conveniently obtained by accessing the route planning service through the application programming interface (API) provided by online map service providers.

$$T_{i,j}^{\text{drive}} = \text{rp}\left(x_{\text{EV},i}, y_{\text{EV},i}, x_{\text{CS},j}, y_{\text{CS},j}\right) \quad (6)$$

where rp(•) denotes the route planning service which outputs the trip duration from the current position of EV *i* $(x_{\text{EV},i}, y_{\text{EV},i})$ to the location of its target station $(x_{\text{CS},j}, y_{\text{CS},j})$. In fact, this is exactly how EV users arrange their trips with map APPs in real life.

Estimated waiting time $T_j^{\text{wait}}$ is calculated and uploaded by CS *j*. Assuming that in CS *j*, the number of EVs queueing and charging are $N_{\text{CS},j}^{\text{EV,queue}}$ and $N_{\text{CS},j}^{\text{EV,ch}}$ respectively, and the remaining charging time of charging EVs are ascending sorted as $(t_1^{\text{remain},j}, \cdots, t_{N_{\text{CS},j}^{\text{EV,ch}}}^{\text{remain},j})$, then $T_j^{\text{wait}}$ is calculated by

$$T_j^{\text{wait}} = \begin{cases} 0, & N_{\text{CS},j}^{\text{EV,ch}} < N_{\text{CS},j}^{\text{EV,max}} \\ t_{1+N_{\text{CS},j}^{\text{EV,queue}}}^{\text{remain},j}, & N_{\text{CS},j}^{\text{EV,ch}} = N_{\text{CS},j}^{\text{EV,max}} \end{cases} \quad (7)$$

*3) Cost integration*

The theory of mental accounting highlights that the perceived value of choices does not strictly align with the utility of those choices [29]. Individuals with bounded rationality establish a reference point to evaluate relative gains and losses, typically exhibiting risk-averse tendencies. Moreover, as one moves away from this reference point, the incremental changes in perceived utility diminish [30]. Consequently, this dynamic results in a skewed value function for users. In a general context, if we denote the utility of an event as *x*, then the perceived value $v(x)$ can be mathematically expressed as follows:

$$v(x) = \begin{cases} (x - x_0)^\alpha, & x > x_0 \\ -\lambda(x_0 - x)^\beta, & x \leq x_0 \end{cases} \quad (8)$$

where parameters $\alpha, \beta$ ($0 < \alpha, \beta < 1$) and $\lambda$ ($\lambda \geq 1$) reflect the user's decision preferences. Fig.4 shows the shape of the value function, where the risk aversion is reflected by $|v(x_0 - x_1)| > |v(x_0 + x_1)|$, and the marginal effect is reflected by $v(x_0 + 2x_1) < 2v(x_0 + x_1)$.

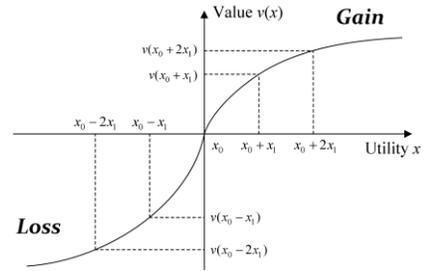

Fig. 4. Value function.

In the context of this charging decision problem, CSs are typically listed in charging APPs in distance-ascending order. This implies that for an EV user, the reference point is the CS with the least time cost. Consequently, any additional time costs of other CSs are perceived as losses. When the expense cost exceeds that of the reference CS, the excess expenses are considered losses; conversely, if the expense cost is lower, the money saved is considered gains. Essentially, the utility derived from time or expense is inversely related to the time or expense cost. Let $j_0^i$ denote the reference CS for EV *i*, then the perceived time value of CS *j* is

$$v_{i,j}^T = -\lambda(-T_{i,j_0^i} + T_{i,j})^\beta \quad (9)$$

since $\forall j \in J, -T_{i,j} \leq -T_{i,j_0^i}$.

The perceived expense value of CS *j* is

$$v_{i,j}^M = \begin{cases} (-M_{i,j} + M_{i,j_0^i})^\alpha, & -M_{i,j} \geq -M_{i,j_0^i} \\ -\lambda(-M_{i,j_0^i} + M_{i,j})^\beta, & -M_{i,j} < -M_{i,j_0^i} \end{cases} \quad (10)$$

Applying the revenue splitting way of value integration in the mental account theory [31], the values of time and expense can be integrated as the comprehensive value $V_{i,j}$ by

$$V_{i,j} = v_{i,j}^M + \theta_i v_{i,j}^T \quad (11)$$

where $\theta_i > 0$ is named as time sensitivity coefficient. The time sensitivity coefficient of different users can be different, since they have different financial conditions and vehicle usages.

*4) Choice criteria*

Generally, EV users are more likely to choose the CSs with greater comprehensive value. But in extreme situations where EV *i* thinks all the CSs have unacceptable prices, it may just choose none of the CSs and delay the charging plan. This choice can also be equivalently seen as "an extra virtual CS indexed -1 and denoted by $j_{-1}$", whose expense cost $M_{i,-1}$ can be seen as that of a "normal" price (e.g. the average price at the time point in a month), and the time cost $T_{i,-1}$ is calculated by

$$T_{i,-1} = T_{i,j_0} + t_i^{\text{limit}} \frac{Soc_{\text{safe}}}{1-Soc_{\text{safe}}} \left( \frac{1}{Soc_i} - 1 \right) \quad (12)$$

where $Soc_{\text{safe}}$ is a safety threshold of SOC, usually set as 0.1. (12) is applied to evaluate the time cost of delaying charging because it satisfies 3 properties: 1) $Soc_i \to 0, T_{i,-1} \to \infty$, meaning EV *i* cannot delay charging since it is about to break down; 2) $Soc_i \to 1, T_{i,-1} \to T_{i,j_0}$, meaning EV *i* can freely delay the charging at no cost when it has adequate electricity; 3) $Soc_i \to Soc_{\text{safe}}, T_{i,-1} \to T_{i,j_0} + t_i^{\text{limit}}$, meaning that the battery is not dying, but cannot sustain subsequent trips, so EV *i* can delay charging at most until the time limit.

To capture the characteristics of the stochastic choice process, the Logit model is applied in the EV choice model [28]. For EV *i*, let $\boldsymbol{J}^+$ denote the set of all CSs and the virtual CS (i.e., $\boldsymbol{J}^+ = \boldsymbol{J} \cup \{j_{-1}\}$), then the choice probability of CS *j* is proportional to the natural logarithm of $V_{i,j}$:

$$\rho_{i,j} = \frac{\exp(V_{i,j})}{\sum_{j \in \boldsymbol{J}^+} \exp(V_{i,j})}, j \in \boldsymbol{J}^+ \quad (13)$$

In summary, the decision-making model is established by (1)-(13). In this model, the uncertainty of EV choices is mainly featured in two aspects:

1. the time sensitivity coefficients of EVs are stochastic;
2. even if the comprehensive values of each CS are known, the final choice of the EV is probabilistic.

The uncertainty makes it impossible for CSs to precisely forecast the choices of EVs. Thus, CSs make pricing decisions to maximize the mathematical expectations of CSAs' profits.

*B. CSAs' decision-making model*

*1) Decision variables*

The pricing is a game among CSAs to maximize their total profit, and the decision variable is the prices of subordinate CSs. Assume that $N^{\text{CS}}$ CSs belong to $N^{\text{CSA}}$ CSAs, the number of CSs in CSA *k* is $N_{\text{CSA},k}^{\text{CS}}$, and the number of EVs proposing charging requests is $N^{\text{EV}}$. Let $\boldsymbol{J}_{\text{CSA},k}$ denote the subordinate CSs of CSA *k*. Hence, the decision variable of CSA *k* is the real-time charging price, denoted by $\{p_j, j \in \boldsymbol{J}_{\text{CSA},k}\}$. The prices do not change for the same EV once it starts charging.

The price adjusting period is set as 5 minutes, primarily for two specific reasons:

1. Prompt Attraction and Influence: A 5-minute pricing period offers a swift interval for CSs to entice potential customers, possibly altering an EV's decision.

2. Alignment with Electricity Market Settlement: Notably, the settlement period within electricity markets, such as 5 minutes in PJM or multiples of 5 minutes in Germany, makes the 5-minute timeframe convenient for CSs. This duration enables CSs to optimize operations while factoring in electricity pricing adjustments effectively.

*2) Objective*

The objective of CSA *k* is to maximize the total profit:

$$\max R_{\text{CSA},k} = R_{\text{CSA},k}^{\text{charge}} + R_{\text{CSA},k}^{\text{DR}} - C_{\text{CSA},k}^{\text{grid}} \quad (14)$$

Public welfare aims are not included in the objective. Nevertheless, demand-side management measures like demand response (DR) can financially encourage CSAs to contribute to public welfare.

The 3 terms are explained in detail as follows:

*1. charging income*

The charging income is the mathematical expectation of expense cost of possible arriving EVs, as (15):

$$R_{\text{CSA},k}^{\text{charge}} = \sum_{i \in \boldsymbol{I}} \sum_{j \in \boldsymbol{J}_{\text{CSA},k}} E_i p_j \sigma_{i,j} \quad (15)$$

where $\sigma_{i,j}$ is the probability that CS *j* is chosen by EV *i*. Note that $\sigma_{i,j}$ is different from $\rho_{i,j}$ in (13), because when calculating $\rho_{i,j}$ in EV *i*'s point of view, $\theta_i$ is regarded as a constant, while in CS *j*'s point of view, $\rho_{i,j}$ is a function of a stochastic parameter $\theta$, denoted by $\rho_{i,j}(\theta)$. If we let $f_{\text{pdf}}(\bullet)$ denote the probability density function of $\theta$, then $\sigma_{i,j}$ is calculated by

$$\sigma_{i,j} = \int_0^{+\infty} \rho_{i,j}(\theta) f_{\text{pdf}}(\theta) d\theta \quad (16)$$

Apparently, (16) is very complex, since $\rho_{i,j}(\theta)$ involves plenty of parameters, making it hard to be dealt with in an analytical form. To ensure the practicability of the proposed method, the distribution of $\theta$, can be approximated by discrete typical values, denoted by $\boldsymbol{\Theta}^{\text{value}} = (\theta_1^{\text{dis}}, \theta_2^{\text{dis}}, \cdots, \theta_Z^{\text{dis}})$, and the corresponding probabilities of these typical values, denoted by

$$\boldsymbol{\Theta}^{\text{prob}} = \left( \int_0^{\theta_1^{\text{dis}}} f_{\text{pdf}}(\theta)d\theta, \int_{\theta_1^{\text{dis}}}^{\theta_2^{\text{dis}}} f_{\text{pdf}}(\theta)d\theta, \cdots, \int_{\theta_{Z-1}^{\text{dis}}}^{\infty} f_{\text{pdf}}(\theta)d\theta \right).$$

Thus, $\sigma_{i,j}$ can be thereby approximated by

$$\sigma_{i,j} = \sum_{z=1}^{Z} \rho_{i,j}\left(\boldsymbol{\Theta}^{\text{value}}(z)\right) \cdot \boldsymbol{\Theta}^{\text{prob}}(z) \quad (17)$$

As we can infer, the greater *Z* is, the more precise (17) will be, while more calculation resource will be necessary.

*2. demand response income*

DR is widely utilized in demand management by temporarily reducing loads. [32] [33].

where $p_0^{\text{DR}}$ is the electricity price during the DR period.

In incentive-based DR, consumers are incentivized to reduce their load, receiving compensation for committing to load reductions. However, they may face penalties if they fail to meet their commitment. A prominent example is the Base Interruptible Program (BIP) administered by the Pacific Gas & Electric Company (PG&E) in California, US [32]. Under this program, users receive an incentive of 8–9$/kW per month. During a DR event, they are charged 6$/kW for any additional demand exceeding the pre-selected demand level (firm service level, FIL). Users are notified of the DR event 30 minutes in advance. Since the monthly incentive is provided regardless of meeting the FIL requirement, in the pricing stage, the primary concern revolves around the penalty charge. Hence, the DR income of CSA $k$ can be expressed as:

$$R_{\text{CSA},k}^{\text{DR}} = r_{\text{CSA},k}^{\text{DR,PRE}} - \omega \max\{q_{\text{CSA},k}^{\max,\text{DR}} - q_{\text{CSA},k}^{\text{FIL}}, 0\} \quad (18)$$

where $r_{\text{CSA},k}^{\text{DR,PRE}}$ is the pre-paid DR incentive, which is a constant in the pricing stage, $\omega$ is the DR penalty price (PP), $q_{\text{CSA},k}^{\max,\text{DR}}$ is the maximal total demand of subordinate CSs during the DR event, and $q_{\text{CSA},k}^{\text{FIL}}$ is the FIL of CSA $k$. Let $t_0$ denote the current time point, if EV $i$ proposed a charging request and will be charged at CS $j$, then the charging power of EV $i$ at CS $j$ at any time point $t \in [t_0, +\infty]$ can be expressed as (19).

$$q_{\text{EV},i}(t) = \begin{cases} 0, & t_0 \le t < t_0 + T_{i,j} \\ q_{i,j}^{\text{ch,max}} & t_0 + T_{i,j} \le t \le t_0 + T_{i,j} + E_{i,j}/q_{i,j}^{\text{ch,max}} \\ 0, & t > t_0 + T_{i,j} + E_{i,j}/q_{i,j}^{\text{ch,max}} \end{cases} \quad (19)$$

Let $\hat{\mathbf{I}}_j$ denote the set of EVs that are charging at CS $j$ at $t_0$, then the charging power of EV $\hat{i} \in \hat{\mathbf{I}}_j$ at any time point $t \in [t_0, +\infty]$ can be expressed as (20)

$$q_{\text{EV},\hat{i}}(t) = \begin{cases} q_{\hat{i},j}^{\text{ch,max}}, & t_0 \le t \le t_0 + t_{\hat{i}}^{\text{remain},j} \\ 0, & t > t_0 + t_{\hat{i}}^{\text{remain},j} \end{cases} \quad (20)$$

where $t_{\hat{i}}^{\text{remain},j}$ is the remaining charging time of EV $\hat{i}$ at CS $j$.

Hence, $q_{\text{CSA},k}^{\max,\text{DR}}$ can be calculated by

$$q_{\text{CSA},k}^{\max,\text{DR}} = \max_t \left\{ \sum_{j \in \mathbf{J}} \left[ \sum_{i \in \mathbf{I}} q_{\text{EV},i}(t)\sigma_{i,j} + \sum_{\hat{i} \in \hat{\mathbf{I}}_j} q_{\text{EV},\hat{i}}(t) \right] \right\} \quad (21)$$

*3. grid electricity cost*

The cost for CSA $k$ to purchase electricity from the power grid is determined by the real-time electricity price.

$$C_{\text{CSA},k}^{\text{grid}} = \sum_{i \in \mathbf{I}} \sum_{j \in \mathbf{J}_{\text{CSA},k}} E_i p_0 \sigma_{i,j} \quad (22)$$

where $p_0$ is the electricity price of the grid.

*3) Constraints*

The price boundary constraint:

$$p_0 \le p_j \le p_{\max}, \forall j \in \mathbf{J}_{\text{CSA},k} \quad (23)$$

where $p_{\max}$ is the upper bound of charging prices, usually set by the government policies.

## IV. EVOLUTIONARY GAME FORMULATION

The pricing strategies adopted by CSAs rely on the expected response from EVs. While this procedure resembles a Stackelberg game, as observed in prior studies, the reality is that the payoff functions and the final decisions of EVs entail uncertainty. Consequently, the pricing process emerges as a single-layer non-cooperative game among CSAs structured as follows:

*Players*: All CSAs.

*Strategies*: Each CSA selects a charging price set for its subordinate CSs.

*Payoffs*: CSAs receive payoffs as depicted in equation (14).

Traditional game theory encounters several challenges in solving this problem due to the following reasons:

*Analytical Complexity*: The derivative of payoff functions cannot be analytically expressed due to the inherent nature of EVs' and CSAs' decision models. This maks it challenging to handle this problem in a continuous action space.

*Limited Rationality*: Assumptions of completely rational players are unrealistic. Players' decisions are shaped by incomplete information and bounded rationality.

*Existence of Equilibrium*: For players utilizing pure strategies within a discretized action space, there's no theorem guaranteeing the existence and uniqueness of a pure strategy Nash Equilibrium (NE) [9]..

Conventional heuristic algorithms like particle swarm optimization (PSO) algorithm (studied in a former paper by the author [36]) or genetic algorithm (GA) also face challenges due to computational resource costs and their inability to consistently find an explainable and reasonable solution.

Given the limitations of prior methodologies, Evolutionary Game (EG) emerges as a viable approach to address imperfect information games. EG implements an imitation of evolutionary processes, gradually filtering out the most fitting strategies, known as Evolutionary Stable Strategies (ESSs) [34][35]. In this paper, an expectation-oriented two-stage EG is formulated to identify ESSs for CSAs. Further details are elucidated below.

### A. Player abstraction

Every CSA is abstracted as a fleet comprising individuals employing diverse strategies within an EG strategy set. Due to computational constraints, this strategy set represents only a fraction of the entire action space. In other words, the decision time is streamlined to a manageable level, at the expense of global optimality.

Individuals within this fleet evaluate their fitness within the game environment and probabilistically switch to other strategies based on their assessed fitness. This iterative process continues until one or more ESSs are identified.

### B. Payoff function

The payoff of CSAs is the profit expectation of the CSA, as shown in (14), which is affected by the strategies adopted by other CSAs. Therefore, in the EG, the fitness in the evolution environment should consider the probabilities of the strategy combination of other CSAs, i.e., "the expectation of the profit expectation", as stated below.

Assume that CSA $k$ has $H_k$ optional strategies, and the $h_k$-th strategy is denoted by $u_{h_k}^k$ ( $h_k = 1, 2, \cdots, H_k$ ). The population share of $u_{h_k}^k$, which is also the probability that $u_{h_k}^k$ is adopted by CSA $k$ in the current iteration stage, is denoted by $x_{h_k}^k$ (by definition, $\sum_{h_k=1}^{H_k} x_{h_k}^k = 1$). Hence, the total number of strategy combinations of all other CSAs, denoted by $H_{k_-}$, is

$$H_{k-} = \prod_{\substack{l=1,\cdots,N_{\text{CSA}} \\ l \neq k}} H_l \tag{24}$$

In the $H_{k-}$ combinations, the $h_{k-}$-th strategy combination is denoted by $u_{h_{k-}}^{k-} = \{u_{h_1}^1, \cdots, u_{h_{k-1}}^{k-1}, u_{h_{k+1}}^{k+1}, \cdots, u_{h_{N_{\text{CSA}}}}^{N_{\text{CSA}}}\}$ ( $h_{k-} = 1, 2, \cdots, H_{k-}$ ). The probability of $u_{h_{k-}}^{k-}$ is

$$x_{h_{k-}}^{k-} = \prod_{\substack{l=1,\cdots,N_{\text{CSA}} \\ l \neq k}} x_{h_l}^l \tag{25}$$

Let $f(u_{h_k}^k, u_{h_{k-}}^{k-})$ denote the payoff when CSA $k$ adopts strategy $u_{h_k}^k$ and other CSAs adopt strategy combination $u_{h_{k-}}^{k-}$. Thus, in the fleet of CSA $k$, the fitness of strategy $u_{h_k}^k$ is

$$F_{h_k}^k = \sum_{h_{k-}=1}^{H_{k-}} f(u_{h_k}^k, u_{h_{k-}}^{k-}) \cdot x_{h_{k-}}^{k-} \tag{26}$$

Furthermore, the average fitness of the fleet of CSA $k$ can be calculated by

$$\overline{F}^k = \sum_{h_k=1}^{H_k} F_{h_k}^k x_{h_k}^k \tag{27}$$

### C. Replicator dynamics

Another important concept in EGs is replicator dynamics, which describes the mechanism that individuals learn from other individuals and transfer to better strategies at a certain rate. In the designed EG of CSAs, the replicator dynamics can be written as:

$$\frac{dx_{h_k}^k}{dt} = x_{h_k}^k \cdot \frac{F_{h_k}^k - \overline{F}^k}{\overline{F}^k} \tag{28}$$

After enough evolution generation, the evolutionary equilibrium is reached when

$$\frac{dx_{h_k}^k}{dt} = 0, \ \forall u_{h_k}^k \tag{29}$$

The strategies with non-zero proportions when the evolutionary equilibrium is reached are called ESSs. Note that the ESS can be more than one. In this circumstance, CSA $k$ can randomly choose any one of the ESSs to reach the highest income expectation.

### D. Overall procedure of EG solution

In EGs, optimal strategies are derived from subsets of the entire action space. The size of the strategy set is constrained by computational limitations; if too large, it demands excessive calculation time, yet if too small, it risks overlooking potentially effective pricing strategies.

To resolve this dilemma, a two-stage EG approach is devised. The lower-layer EG is to ascertain an approximate range of optimal prices. Subsequently, the upper-layer EG generates ESSs by employing randomly generated strategies within this approximate price range. The detailed procedural steps can be articulated using pseudocode as follows:

---

**Procedure 1: Lower-layer EG**

Input: Optional prices and public information of EVs and CSs.
Output: The ESSs of CSAs.

// *Step 1.* Strategy sets construction
1: Divide the optional prices into $H_k$ groups by equal increment
2: $\{\mathbf{S_k}, k=1,\cdots,N^{CSA}\} = \varnothing$ //Initialize the strategy sets of CSAs
3: for $k = 1$ to $N^{CSA}$ do
4:   for $h_k = 1$ to $H_k$ do
5:     $\mathbf{S_k} \leftarrow \mathbf{S_k} \cup \{N_{CSA,k}^{CS} \times 1$ random values in the $h_k$-th group$\}$ //Add a random strategy within every price group
6:   end for
7: end for
8: do
9:   for $k = 1$ to $N^{CSA}$ do
// *Step 2.* Fitness calculation
10:     Calculate $f(u_{h_k}^k, u_{h_{k-}}^{k-})$ by (15) //The payoffs of all CSAs under every strategy combination
11:     Calculate $F_{h_k}^k$ by (28) //The fitness of strategy $u_{h_k}^k$
12:     Calculate $\overline{F}^k$ by (29) //The average fitness of CSA $k$
// *Step 3.* Population share update
13:     Update $x_{h_k}^k$ by (30) //The population share of strategies
14:   end for
15: until (31) is met
16: return $\{ESS_k^{\text{lower-layer}} = u_{h_k}^k, k=1,\cdots,N^{CSA} \mid h_k : x_{h_k}^k > 0\}$
// *Step 4.* Return he ESSs of CSAs in the lower-layer EG

---

**Procedure 2: Upper-layer EG**

Input: The ESSs in the lower-layer EG and public information of EVs and CSs.
Output: The ESSs of CSAs.

// *Step 1.* Strategy sets construction
1: $\{\mathbf{S_k}, k=1,\cdots,N^{CSA}\} = \varnothing$ //Initialize the strategy sets of CSAs
2: for $k = 1$ to $N^{CSA}$ do
3:   for $h_k = 1$ to $H_k$ do
4:     $\mathbf{S_k} \leftarrow \mathbf{S_k} \cup \{ESS_k^{\text{lower-layer}} + N_{CSA,k}^{CS} \times 1$ random values within the deviation range$\}$
5:   end for
6: end for
7: do
8:   for $k = 1$ to $N^{CSA}$ do
// *Step 2.* Fitness calculation
9:     Calculate $f(u_{h_k}^k, u_{h_{k-}}^{k-})$ by (15) //The payoffs of CSAs under every strategy combination
10:     Calculate $F_{h_k}^k$ by (28) //The fitness of strategy $u_{h_k}^k$
11:     Calculate $\overline{F}^k$ by (29) //The average fitness of CSA $k$
// *Step 3.* Population share update
12:     Update $x_{h_k}^k$ by (30) //The population share of strategies
13:   end for
14: until (31) is met
15: return $\{ESS_k^{\text{upper-layer}} = u_{h_k}^k, k=1,\cdots,N^{CSA} \mid h_k : x_{h_k}^k > 0\}$
// *Step 4.* Return the ESSs of CSAs in the upper-layer EG

---

As clarified above, the evolutionary equilibrium is different from the Nash equilibrium. Therefore, an indicator is needed for evaluating the performance of the strategy. In [19], the authors use the No-regret Index (NI) to measure the proximity of the solution to NE. Similarly, the NI of this problem is defined as the ratio of CSAs' payoffs to the payoffs of the best-response action in the whole action space $S_{AS}^k$ when other CSAs retain their current strategies:

$$NI = \frac{\sum_{k=1}^{N^{CSA}} f(u_{h_k}^k, u_{h_{k-}}^{k-})}{\sum_{k=1}^{N^{CSA}} \max_{u_{h_k}^k \in S_{AS}^k} f(u_{h_k}^k, u_{h_{k-}}^{k-})} \quad (30)$$

The closer NI is to 1, the better the obtained solution is. If NI = 1, then the NE is reached. In [19], $\max_{u_{h_k}^k \in S_{AS}^k} f(u_{h_k}^k, u_{h_{k-}}^{k-})$ is calculated by enumerating. However, in this problem, $S_{AS}^k$ is too large for enumerating. Therefore, a modified NI, named No-regret Index in Sampled-action-space (NIS), is adopted for performance evaluation in this paper. The difference between NIS and NI is that the whole action space $S_{AS}^k$ is replaced by a sampled action space $S_{SAS}^k$, which is randomly sampled from the whole action space, and is enumerable. I.e.,

$$NIS = \frac{\sum_{k=1}^{N^{CSA}} f(u_{h_k}^k, u_{h_{k-}}^{k-})}{\sum_{k=1}^{N^{CSA}} \max_{u_{h_k}^k \in S_{SAS}^k} f(u_{h_k}^k, u_{h_{k-}}^{k-})} \quad (31)$$

*E. Difference of pricing by CSAs and CSs*

Table III summarizes the primary distinctions between pricing as CSAs and pricing as individual CSs based on the decision model and the proposed solution method. Strategy sampling noticeably reduces computational complexity, at the cost of full optimality. In conclusion, adopting the form of CSAs for pricing provides considerable advantages in terms of model rationality and solvability.

TABLE III
DIFFERENCE BETWEEN PRICING AS CSAS AND AS CSS

| Aspect | | Difference | Comments |
|---|---|---|---|
| Decision variable | CSA | $N_{CSA,k}^{CS} \times 1$ vector $\{p_j, j \in \mathbf{J_{CSA,k}}\}$ | Dimensional difference in decision variable |
| | CS | Scalar $p_j$ | |
| Number of players | CSA | $N_{CSA}$ | $N_{CSA} < N_{CS}$ |
| | CS | $N_{CS}$ | |
| Objective | CSA | Maximal combined profit of subordinate CSs | Pricing as CSs brings meaningless internal competition within the CSA, lowering the overall profit |
| | CS | Maximal profit of itself | |
| Problem solving difficulty | CSA | Requires calculating payoffs in $\prod_{k \in \mathbf{K}} H_k$ strategy combination scenarios | Pricing in the form of CS requires much more calculation resources, making it hard to meet real-time pricing requirement |
| | CS | Requires calculating payoffs of every CS in $\prod_{j \in \mathbf{J}} H_j$ strategy combination scenarios | |

## V. CASE STUDY

*A. Case introduction*

To verify the framework and methodology of this paper, a typical area in the Pudong District, Shanghai, China, is considered. 83 CSs are located in this area, of which 28 belong to Star Charge, 46 belong to TELD, and 9 belong to State Grid. Naturally, the 3 CSPs play the role 3 CSAs. The randomly generated 300 charging requests are based on a charge load forecasting module developed by the author team [37]. Fig. 5 gives a brief overview of the case area, with the CSs and charging requests labeled on the map.

The upper bound of the charging price is set at 3.0 ¥/kWh, and the electricity price is set as 0.5 ¥/kWh. Therefore, the optional prices range from 0.5 to 3.0 in increments of 0.1, totaling 26. The time sensitivity coefficient is assumed to obey the distribution $\theta \sim N(0.1, 0.025^2)$. The digital map service used in this case study is provided by Baidu Map [38], which is one of the most popular online map services in China. According to existing studies [39][40][41], the parameters of the value function and weight function can be set as $\alpha = \beta = 0.88$, $\lambda = 2.25$.

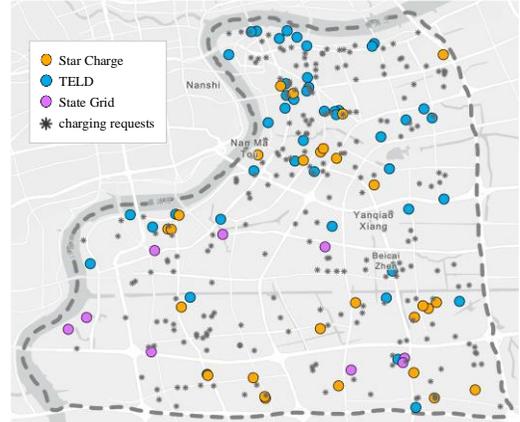

Fig. 5 Overview of the case area.

In the simulation, real-time positions of EVs are calculated according to the direction route given by the digital map API, while in practice they are automatically uploaded under the proposed framework. Besides this, all information is generated and interacted in the same way as it is in reality, which verifies the practicability of the proposed framework.

*B. Base scenario*

In the base scenario, we assume that no DR is implemented, and CSAs have a precise grasp of the $\theta$ distribution of EVs.

The simulation is implemented in MATLAB R2021b, on a laptop with an Intel i7-1165G7 @ 2.80GHz CPU. The running time of the lower-layer and upper-layer EG are 10.27s and 16.65s, respectively. Since the total running time is 26.92s, the instantaneity of the proposed method is capable of real-time pricing in the real-world, which guarantees the practicability.

The EG strategy set of each CSA comprises 13 strategies in the lower-layer EG, which are randomly generated within the 26 optional prices, with a 0.2 increment segmentation. The ESSs in the lower-layer EG are modified to derive 16 strategies for each CSA in the upper-layer EG. As shown in Fig. 6, the population share of strategies of the 3 CSAs are initialized equal, and one of the strategies emerge from them, whose population share approached 1 within 500 iteration generations.

Along with the variation of the strategy population shares, the average fitness (i.e., profit expectation) also varies in the process, as shown in Fig.7.

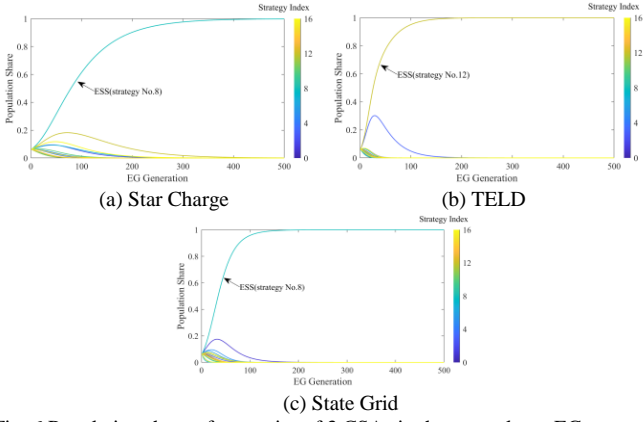

Fig. 6 Population share of strategies of 3 CSAs in the upper-layer EG.

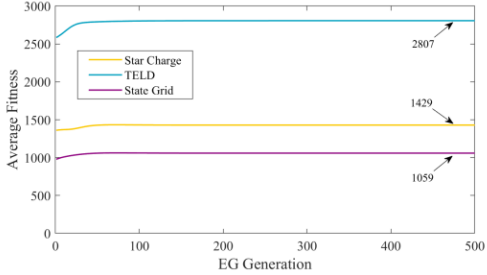

Fig. 7 Average fitness of 3 CSAs in the upper-layer EG.

The decided prices of CSAs are shown in Fig.8. The bars in orange, blue and purple are the charging prices of Star Charge, TELD and State Grid respectively. As shown in Fig. 8, the prices are distributed in range [0.7, 1.2]. This is slightly lower than that in reality. By our inference, this mainly attributes to: 1) Operation costs other than electricity expenses are not considered; 2) The pricing in the proposed framework is under a highly competitive environment, while in real-world it is not.

As mentioned in section IV. D, the prices are evaluated with NIS, as illustrated in Fig. 9. In Fig.9, the dashed lines labeled ①, ②, ③ are the profit expectations of Star Charge, TELD and State Grid under the equilibrium. The scatters represent the sampled action space of CSAs, formed by 1000 randomly sampled strategies respectively.

The scatters in the orange area (indexed 1-1000) show the profit expectations when Star Charge adopts strategies in the sampled action space, while TELD and State Grid retain their ESSs. The 1000 orange scatters in this area are all beneath line ①, which means that for Star Charge, all the strategies in the sampled action space have worse performance than its ESS, i.e., $f(u_{h_1}^1, u_{h_{1-}}^{1-}) = \max_{u_{h_1}^1 \in S_{sampled}^1} f(u_{h_1}^1, u_{h_{1-}}^{1-})$. Meanwhile, the other two CSAs will benefit from the strategy deviation of Star Charge.

Similarly, the blue area (index 1001-2000) and purple area (index 2001-3000) illustrate the optimality of the ESSs in the sampled action space, i.e., $f(u_{h_2}^2, u_{h_{2-}}^{2-}) = \max_{u_{h_2}^2 \in S_{sampled}^2} f(u_{h_2}^2, u_{h_{2-}}^{2-})$, $f(u_{h_3}^3, u_{h_{3-}}^{3-}) = \max_{u_{h_3}^3 \in S_{sampled}^3} f(u_{h_3}^3, u_{h_{3-}}^{3-})$. Hence, according to (31), NIS=1. This indicates that the proposed two-stage EG method is capable of finding very good pricing strategies for all game players (although they are probably not the optimal ones).

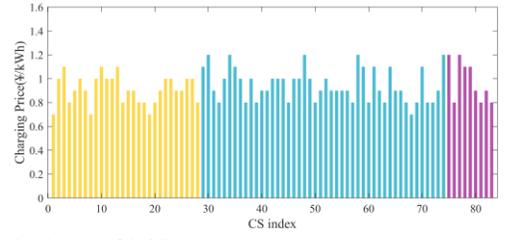

Fig. 8 Decided price of 3 CSAs.

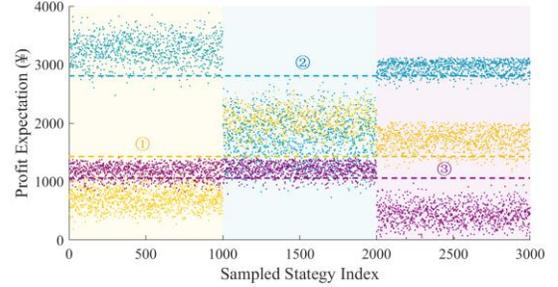

Fig. 9 Profit expectation of sampled strategies.

### C. Charging demand quantity effects

In the base scenario, 300 charging requests are distributed in the case area. In this scenario, the charging demand is multiplied by coefficients 0.5, 1, 2, 4 to simulate different charging demand quantity conditions.

As shown in Fig.10, the average charging price decreases as the charging demands increase. This result reveals the benefit of dynamic pricing: full market competition helps reducing the charging cost of EV users. The fiercer CSAs' competition is, the lower the average price is.

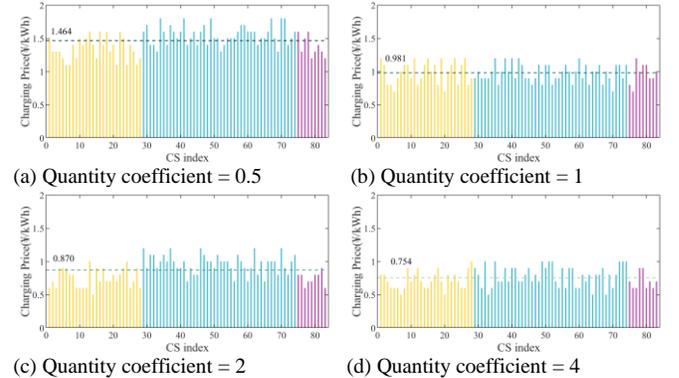

Fig. 10 Charging prices under charging demand quantity coefficient = 0.5, 1, 2, 4.

### D. Charging request position effects

The geographic positions of EVs will affect time costs and therefore act on the pricing of CSs. An extreme contrast scenario as shown in Fig.11 (a) is designed to visualize this effect. In this contrast scenario, 300 EVs are concentrated in the north-west corner of the area with other parameters unchanged. The colored circles denote CSs' charging prices, variating from 0.7 to 1.4 and colored in green to red. Intuitively, red circles are more concentrated in the north of the area. Divided by Longyang road, the average price of the CSs in the north, where EV charging requests are concentrated, is 1.146 ¥, which is 13.80% higher comparing to the south. In contrast, in the base scenario, as illustrated in Fig. 11 (b), 300 EVs are evenly distributed, and the charging prices between the north and the south have no obvious difference (-1.52%).

The comparation between this two scenarios reveals that the location advantage attributes locational price premium to the charging prices. This also indicates the value of information on the geographic position of EVs for CS pricing.

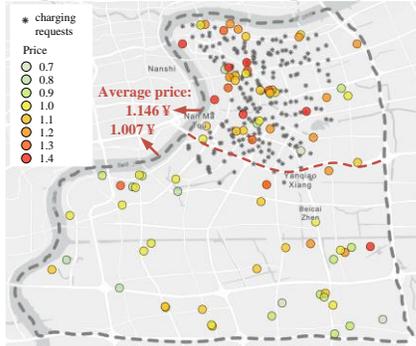

(a) North-west concentrating scenario

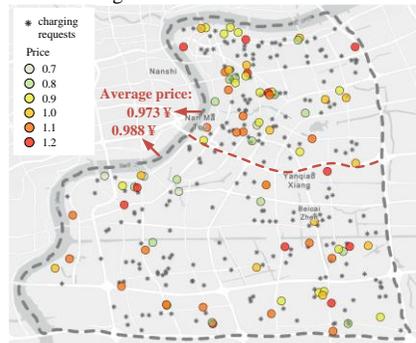

(b) Base scenario
Fig. 11 Charging prices under the north-west concentrating scenario and the base scenario.

### E. $\theta$-sensitivity analysis

As stated in section III, $\theta$-distribution is a significant reference for the decision of CSAs. To reveal the effects of the perceived $\theta$ distribution, a sensitivity analysis is conducted.

In the base scenario, it is assumed that $\theta \sim N(0.1, 0.025^2)$, i.e., $\mu = 0.1$. As a contrast, in this analysis, we assume that Star Charge mistakenly perceive $\mu$ as 0.2, 0.4, 0.8, and makes pricing decisions accordingly, while TELD and State Grid correctly perceive $\mu$ as 0.1.

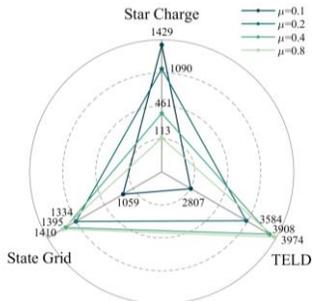

Fig. 12 Profit expectation under $\mu = 0.1, 0.2, 0.4, 0.8$.

The radar chart in Fig. 12 shows the profit expectation of the 3 CSAs under the four scenarios of $\mu = 0.1, 0.2, 0.4, 0.8$. As can be observed, the profit expectation drops while the perceived $\mu$ by Star Charge deviates farther from the reality. Meanwhile, competitors benefit from the mistake of Star Charge.

This result reveals the data asset attribute of $\theta$-distribution. Since this information directly affects the profit of CSAs, CSAs will attempt to grasp a precise profile of distribution. Therefore, it is natural to regard it as private information, which is shared only within the CSA itself.

### F. DR effect analysis

As stated in section III, if DR is to be implemented, the DR income should be considered in the pricing. In this sub-section, the effect of different DR conditions is analyzed. In incentive DRs, FIL (firm service level) and PP (penalty price) are the key conditions. The basic DR scenario is set as FIL = (28, 46, 9) / (28 + 4 + 9) × 60 kW/h × 300 = (6072.3, 9975.9, 1951.8) kW for the 3 CSAs, and PP = 6$/kW.

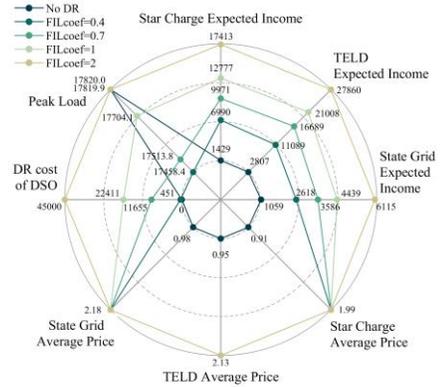

(a) Metrics under FIL coefficient = 0.4, 0.7, 1, 2

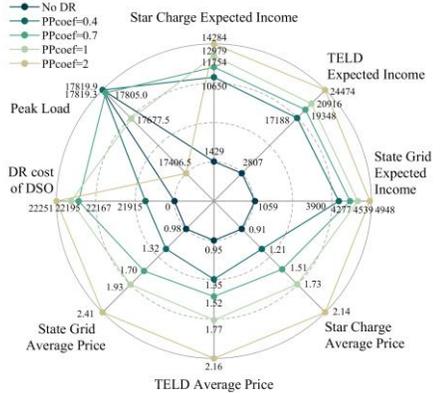

(b) Metrics under PP coefficient = 0.4, 0.7, 1, 2
Fig. 13 Metrics under different DR conditions.

Fig. 13 (a) shows the metrics under scenarios when the basic FIL is multiplied by coefficients 0.4, 0.7, 1, 2. As can be observed, the profit expectation of 3 CSAs increase with the FIL, which is mainly because the pre-paid DR incentive is proportional to FIL. A counterintuitive finding is that FIL has no significant effects on the prices. The peak load of the whole region at FIL coefficient = 2 is the same as that when no DR is implemented, but peak shaves are observed at FIL coefficient = 1 or less. The DR cost of DSO at FIL coefficient = 2 is approximately double that at FIL coefficient = 1, but when the FIL coefficient is 0.7 or 0.4, the DR cost is less than the proportion. As we can infer, the FIL restriction forces CSAs to pay the fine. This will foreseeably influence the intention for signing the DR contract, although quantitative analysis on this topic is not conducted since it is out of the focus of this paper. In summary, too high or too low FIL should both be avoided.

Fig. 11 (b) shows the metrics under scenarios when the basic PP is multiplied by coefficients 0.4, 0.7, 1, 2. The profit expectations and average prices of the 3 CSAs increase as the penalty price increases. The peak load is hardly shaved at PP coefficient = 0.4 and 0.7 compared to the no-DR scenario, which illustrated that too low PP will weaken the effectiveness of DR. In this case study, the peak load decreases significantly when PP coefficient increases from 1 to 2, with little DR cost increase. It can be seen that increasing PP is an economical and effective measure to control the peak load. However, similar to the discussion on FIL above, this may also require a trade-off between DR effectiveness and contracting intention.

## VI. Conclusion

This paper mainly focuses on the dynamic pricing of EV charging station alliances under a bidirectional information interaction framework. The result of the case study based on a real area in Shanghai with 83 CSs verifies the practicability of the framework and the feasibility of the pricing strategy. The proposed model effectively considered the bounded rationality of EV users and the information asymmetry between EVs and CSs. The proposed two-stage EG shows outstanding performance in finding acceptably optimal pricing strategies within acceptable solution time. The application of online map service API offers an efficient and precise method for road network modeling. Future studies may focus on the influence of government policies on the pricing of CSs.